\documentstyle[12pt,a4,epsf]{article}
\title{Radiative Rare $B$ Decays \\
       into Higher $K$-Resonances}
\author{\hfil\\
        {\sc Ahmed Ali},\quad
        {\sc Thorsten Ohl}%
                \thanks{Supported by
                a grant of Deutsche Forschungsgemeinschaft.}
                \thanks{Address after Sept.~1, 1992: Lyman
                Laboratory of Physics, Harvard University,
                Cambridge, MA 02138, USA.}\\
      \hfil\\
      Deutsches Elektronen-Synchrotron DESY \\
      W-2000 Hamburg, Federal Republic of Germany
      \hfil\\\hfil\\
      {\sc Thomas Mannel}\\
      \hfil\\
      Technische Hochschule Darmstadt, \\
      W-6100 Darmstadt, Federal Republic of Germany
}

\begin{document}
\maketitle
\vfill

%%%%%%%%%%%%%%%%%%%%%%%%%%%%%%%%%%%%%%%%%%%%%%%%%%%%%%%%%%%%%%%%%%%%%%%%

\begin{abstract}
\noindent
We estimate the contributions of higher $K$-resonances
to the radiative rare decays $b \to s \gamma$.  We use the spin
symmetry for heavy quarks to reduce the number of independent
form factors to four Isgur-Wise functions, which are estimated from the
wave function model of Isgur, Scora, Grinstein, and Wise.  Our results
suggest a substantially larger branching fraction for the decay
$B\to K^*_2(1430)\gamma$ than previous investigations.
\end{abstract}

%%% if not DAHEP %%%%%%%%%%%%%%%%%%%%%%%%%%%%%%%%%%%%%%%%%%%%%%%%%%%%%%%
\vbox to 0pt{\vss
  \vbox to \textheight{\noindent
    DESY 92-113 \hfill ISSN 0418-9833\\
    IKDA 92/23 \hfill\\
    August 1992 \hfill
    \vfill
  }
}
%%% fi %%%%%%%%%%%%%%%%%%%%%%%%%%%%%%%%%%%%%%%%%%%%%%%%%%%%%%%%%%%%%%%%%

\newpage

%%% if DAHEP %%%%%%%%%%%%%%%%%%%%%%%%%%%%%%%%%%%%%%%%%%%%%%%%%%%%%%%%%%%
% \starttext
% \pagestyle{columns}
% \pagenumbering{arabic}
%%% fi %%%%%%%%%%%%%%%%%%%%%%%%%%%%%%%%%%%%%%%%%%%%%%%%%%%%%%%%%%%%%%%%%

%%%%%%%%%%%%%%%%%%%%%%%%%%%%%%%%%%%%%%%%%%%%%%%%%%%%%%%%%%%%%%%%%%%%%%%%

\def\ket#1{\left\langle#1\right\vert}
\def\bra#1{\left\vert#1\right\rangle}
\def\ME#1#2#3{\left\langle#1\left\vert#2\right\vert#3\right\rangle}
\def\slash#1{#1 \hskip -0.5em / }
%\baselineskip 8mm

%%%%%%%%%%%%%%%%%%%%%%%%%%%%%%%%%%%%%%%%%%%%%%%%%%%%%%%%%%%%%%%%%%%%%%%%

\section{Introduction}
\label{sec:intro}

The heavy quark limit has proven to be a useful tool to obtain model
independent information on systems containing heavy
quarks~\cite{IW91,Bjo90,Geo90}.  In the limit~$m_Q \to \infty$
($m_Q$ being
the mass of the heavy quark) two additional symmetries beyond those of
QCD arise. The first one is the so called heavy flavor symmetry, the
masses of the heavy quarks may be scaled out and the lagrangian is the
same for all heavy flavors. Thus there is an $SU(N_f)$ symmetry of
rotations among the heavy quarks. The second symmetry is the spin
symmetry.  In the limit of infinitely heavy quarks the spin degrees of
freedom of the heavy quark decouple and the $SU(2)$ rotations of the
heavy quark spin become a symmetry.  This means that the
spectrum of heavy hadrons should show degenerate spin symmetry
doublets; for the case of heavy mesons built of a heavy quark and a
light antiquark the pseudoscalar ($0^-$) and the corresponding vector
mesons ($1^-$) form such a spin symmetry doublet.

These additional symmetries allow many interesting predictions.
In particular, they imply model independent relations
between form factors for weak decays. For instance, the
semileptonic $B$ decays into the lowest spin symmetry doublet
of $D$-mesons (i.e. the $D$ and $D^*$) are described
in terms of only one independent form factor; in addition
one obtains a model independent statement about the
normalization of this form factor at maximum momentum
transfer.
Also the excited states of heavy hadrons fall into spin symmetry
doublets in the heavy quark limit~\cite{IW91b}.  The corresponding
group theory has
been elaborated in~\cite{Fal92}.  The main result is that for
mesonic transitions from a heavy $0^-$ ground state meson to a heavy
excited meson only one independent form factor per spin
symmetry doublet appears.

Heavy to heavy transitions of a $0^-$-meson into excited states
have been considered in~\cite{IW91b,BHKT92} for the case of
semileptonic $B\to D^{**}$ decays.  Although the $s$-quark is not
particularly heavy and very substantial corrections to the Isgur-Wise
functions are to be anticipated, particularly at the symmetry point,
yet we feel that the heavy quark symmetry can be used gainfully to
relate various form factors.  The
heavy quark limit has been applied to some exclusive rare $b\to s$
decays~\cite{AM91} in order to obtain an estimate of the corresponding
branching fractions, where it has been determined that the branching
ratios depend on the extrapolation of the Isgur-Wise function
$\xi(vv')$ into the region of interest $vv' \gg 1$.  This requires
a reasonable model for the wave functions, for which we take the
Grinstein-Isgur-Scora-Wise (GISW) model~\cite{GISW89}.
In the present paper we extend the work
of~\cite{AM91} to the higher resonances in the $K$ system.  We shall
focus on the two body decays $B\to K^*\gamma$ and
$B\to K^{**}\gamma$ since they are
expected to have larger branching fractions than the corresponding
short distance contributions to the leptonic decays $B\to
K^{**}\ell^+\ell^-$.

A rich spectrum of states has been observed in the $K$
system. In table~\ref{tab:K-resonances} we summarize the present
knowledge about the excited $K$-mesons~\cite{PDG92}.
We use the standard notation for the various quantum numbers. The radial
excitation quantum number is denoted by $n=1,2,3, \cdots$.
$L=(S=0),(P=1),(D=2),\cdots$ is the orbital angular momentum,
$s=0,1$ is the sum of the spins of the two quarks in the meson and
finally $J=1,2,3, \cdots$ is the total spin of the meson. The parity
is given by the orbital angular momentum to be $P = (-1)^{L+1}$.

\begin{table}
  \begin{minipage}{\linewidth}
  {%%% The TeXbook, ch. 22:
   \newdimen\digitwidth\setbox0=\hbox{\rm0}\digitwidth=\wd0
   \catcode`?=\active\def?{\kern\digitwidth}
   \newdimen\dotwidth\setbox0=\hbox{\rm.}\dotwidth=\wd0
   \catcode`,=\active\def,{\kern\dotwidth}
  \begin{center}
  \begin{tabular}{|l|l|c|c|c|c@{}r@{$\pm$}r|}
  \hline
    \multicolumn{1}{|c|}{Name}
      &\multicolumn{1}{|c|}{State}
      &$ J^P $&$ n^{2s+1} L_J  $&$ [j_{light}] $
      &\multicolumn{3}{|c|}{Mass/MeV} \\
  \hline\hline
  $ K           $&$ C      $&$ 0^- $&$ 1^1S_0        $&$ [1/2] $
      & &  497.67 &  0.03 \\
  $ K^*(892)    $&$ C^*    $&$ 1^- $&$ 1^3S_1        $&$ [1/2] $
      & &  896.1? &  0.3? \\
  $ K_1(1270)   $&$ E^*    $&$ 1^+ $&$ 1^1P_1/1^3P_1 $&$ [1/2] $
      & & 1270,?? & 10,?? \\
  $ K_1(1400)   $&$ F      $&$ 1^+ $&$ 1^1P_1/1^3P_1 $&$ [3/2] $
      & & 1402,?? &  7,?? \\
  $ K^*(1410)   $&$ C^*_2  $&$ 1^- $&$ 2^3S_1        $&$ [1/2] $
      & & 1412,?? & 12,?? \\
  $ K^*(1430)   $&$ E      $&$ 0^+ $&$ 1^3P_0        $&$ [1/2] $
      & & 1429,?? &  7,?? \\
  $ K^*_2(1430) $&$ F^*    $&$ 2^+ $&$ 1^3P_2        $&$ [3/2] $
      & & 1425.4? &  1.3? \\
  $ K(1460)     $&$ C_2    $&$ 0^- $&$ 2^1S_0        $&$ [1/2] $
      &$\approx$&\multicolumn{2}{@{}l|}{1460} \\
  $ K_2(1580)   $&$ G^*    $&$ 2^- $&$ [1^1D_2/1^3D_2] $&$ [3/2] $
      &$\approx$&\multicolumn{2}{@{}l|}{1580} \\
  $ K_1(1650)   $&$        $&$ 1^+ $&$ [2^1P_1]      $&$ [1/2] $
      & & 1650,?? & 50,?? \\
  $ K^*(1680)   $&$ G      $&$ 1^- $&$ 1^3D_1        $&$ [3/2] $
      & & 1714,?? & 20,?? \\
  $ K_2(1770)   $&$        $&$ 2^- $&$ [1^1D_2/]1^3D_2 $&$ [5/2] $
      & & 1768,?? & 14,?? \\
  $ K_3^*(1780) $&$        $&$ 3^- $&$ 1^3D_3        $&$ [5/2] $
      & & 1770,?? & 10,?? \\
  $ K(1830)     $&$        $&$ 0^- $&$ 3^1S_0        $&$ [1/2] $
      &$\approx$&\multicolumn{2}{@{}l|}{1830} \\
  $ K^*_0(1950) $&$        $&$ 0^+ $&$ [2^3P_0]      $&$ [1/2] $
      & & 1945,?? & 32,?? \\
  $ K^*_2(1980) $&$        $&$ 2^+ $&$               $&$     $
      & & 1975,?? & 22,?? \\
  $ K_4^*(2045) $&$        $&$ 4^+ $&$ 1^3 F_4       $&$     $
      & & 2045,?? &  9,?? \\
  $ K_2(2250)   $&$        $&$ 2^- $&$               $&$     $
      & & 2247,?? & 17,?? \\
  $ K_3(2320)   $&$        $&$ 3+  $&$               $&$     $
      & & 2324,?? & 24,?? \\
  $ K^*_5(2380) $&$        $&$ 5^- $&$               $&$     $
      & & 2382,?? & 24,?? \\
  $ K_4(2500)   $&$        $&$ 4^- $&$               $&$     $
      & & 2490,?? & 20,?? \\
    \hline
\end{tabular}
\end{center}}
\end{minipage}
\caption{Spectrum of excited neutral $K$-mesons.  Notation, masses, and
         assignment of quantum numbers are taken
         from~\protect\cite{PDG92}, except for the quantities in brackets.
         The systematical and statistical errors on the masses have
         been added in quadrature.}
\label{tab:K-resonances}
\end{table}

The paper is organized as follows.  After a brief review of the spin
symmetry formalism for higher resonances we shall assign spin symmetry
doublets to the states listed in table~\ref{tab:K-resonances}. Based
on the effective Hamiltonian from~\cite{effective Hamiltonian}, we
calculate the
resulting branching ratios in section~\ref{sec:penguins}.  The
undetermined Isgur-Wise functions will be estimated from the GISW wave
function model~\cite{GISW89} in section~\ref{sec:wave-func}.
Finally we compare our estimates with a previous prediction from a
wave function model~\cite{Alt88}, which does not take into account the
spin symmetry.

%%%%%%%%%%%%%%%%%%%%%%%%%%%%%%%%%%%%%%%%%%%%%%%%%%%%%%%%%%%%%%%%%%%%%%%%

\section{Matrix Representation of the States}
\label{sec:tensor}

Heavy quark symmetry will lead to relations between form factors which
may be formulated in terms of a Wigner Eckart theorem.
The calculation of the relevant Clebsch-Gordan Coefficients and the
counting of the number of independent form factors is conveniently
done in the framework of the trace formalism which was formulated
in~\cite{Bjo90,Geo91,KS88} and generalized to excited states
in~\cite{Fal92}.

Following~\cite{Fal92} we write for the lowest lying states
\begin{eqnarray} \label{C}
C(v) &=& \frac{1}{2} \sqrt{m} (\slash{v} + 1) \gamma_5
, \; J^P = 0^-
      \\
C^*(v,\epsilon) &=& \frac{1}{2} \sqrt{m} (\slash{v}+1) \slash{\epsilon}
, \; J^P = 1^-
      \\
E(v) &=& \frac{1}{2} \sqrt{m} (\slash{v} + 1)
, \; J^P = 0^+
      \\
E^* (v,\epsilon) &=&
\frac{1}{2} \sqrt{m} (\slash{v} + 1) \gamma_5 \slash{\epsilon}
, \; J^P = 1^+, j_{light} = 1/2
      \\ \label{F}
F (v,\epsilon) &=& \frac{1}{2} \sqrt{m}
\sqrt{\frac{3}{2}}  (\slash{v} + 1)
\gamma_5 \left[ \epsilon^\mu - \frac{1}{3}
\slash{\epsilon} (\gamma^\mu - v^\mu ) \right]
, \; J^P = 1^+, j_{light} = 3/2
      \\
F^* (v,\epsilon) &=&
\frac{1}{2} \sqrt{m} (\slash{v} + 1) \gamma_\nu \epsilon^{\mu\nu}
, \; J^P = 2^+, j_{light} = 3/2
      \\
G (v,\epsilon) &=& \frac{1}{2} \sqrt{m}
\sqrt{\frac{3}{2}}  (\slash{v} + 1)
\left[ \epsilon^\mu - \frac{1}{3}
\slash{\epsilon} (\gamma^\mu + v^\mu ) \right]
, \; J^P = 1^-, j_{light} = 3/2
      \\ \label{G*}
G^* (v,\epsilon) &=&
\frac{1}{2} \sqrt{m} (\slash{v} + 1) \gamma_5 \gamma_\nu
\epsilon^{\mu\nu}
, \; J^P = 2^-, j_{light} = 3/2
\end{eqnarray}
Here $m$ and $v$ are the mass and the velocity of the heavy meson,
$j_{light}$ denotes the total spin of the light degrees of
freedom and the tensor $\epsilon^{\mu \nu}$ is the polarization
of a spin 2 object with
$\epsilon^{\mu \nu} = \epsilon^{\nu \mu}$ and
$\epsilon^{\mu \nu} v_\nu = 0$.
These eight states fall into four spin symmetry doublets.
The two ground state meson $C (v)$ and $C^* (v, \epsilon)$ form
the first one, $E(v)$ and $E^* (v,\epsilon)$ the second,
$F (v,\epsilon)$ together with
$F^* (v,\epsilon)$ is the third, and $G(v, \epsilon)$ forms the fourth
with $G^*(v,\epsilon)$.

Matrix elements of bilinear currents of two heavy quarks are
calculated by taking the trace:
\begin{equation}
\ME{{\cal H}^\prime (v^\prime)}{\bar{h}^\prime_{v^\prime}
\Gamma h_v^\prime}{{\cal H} (v)}  =
\mbox{Tr} \left\{
{\cal H}^\prime (v^\prime) \Gamma {\cal H} (v)
                {\cal M} (v, v^\prime) \right\}
\end{equation}
where ${\cal H}$ denotes any of the representation
matrices~(\ref{C}-\ref{G*}).
The matrix ${\cal M}$ represents the light
degrees of freedom. As discussed in~\cite{Fal92} this matrix may
be expressed in terms of only one independent form factor for
each spin symmetry doublet. We define the Isgur-Wise functions
for the transitions of a $0^-$ ground state meson into an excited
meson by
\begin{eqnarray}
{\cal M} (v, v^\prime) &=& \xi_C (v  v^\prime ) \hphantom{v_\mu}
  \qquad\mbox{for}\quad 0^- \to (0^- , 1^-) = (C,C^*)       \\
{\cal M} (v, v^\prime) &=& \xi_E (v  v^\prime ) \hphantom{v_\mu}
  \qquad\mbox{for}\quad 0^- \to (0^+ , 1^+) = (E,E^*)       \\
{\cal M} (v, v^\prime) &=& \xi_F (v  v^\prime ) v_\mu
  \qquad\mbox{for}\quad 0^- \to (1^+ , 2^+) = (F,F^*)       \\
{\cal M} (v, v^\prime) &=& \xi_G (v  v^\prime ) v_\mu
  \qquad\mbox{for}\quad 0^- \to (1^- , 2^-) = (G,G^*)
\end{eqnarray}
where the vector index in the last two equations will be contracted
with the one in the representations of the excited mesons,
c.f.~eqs.~(\ref{F})-(\ref{G*}). Note that $\xi_C = \xi$ is the usual
Isgur-Wise function with $\xi (v  v^\prime =1 ) = 1$. We shall also
consider a decay into a radially excited $K$-meson which has the same
spin parity quantum numbers as the ground state. The corresponding
Isgur-Wise function will be denoted by  $ \xi_{C_2}$.

Only for the lowest lying spin symmetry doublet the normalization
of the corresponding Isgur-Wise function is known. The value at the
normalization point $vv^\prime = 1$  of the
Isgur-Wise functions for the higher spin
symmetry doublets may be related to
the slope of the Isgur-Wise function of the lowest doublet using
the Bjorken sum  rule techniques~\cite{Bjo90,BDT91}. However, the
applications discussed below involve the Isgur-Wise functions at
values of $vv'$ much in excess of 1, and hence the information at and
near the normalization point are not very useful in the present context.

We shall discuss rare $B$ decays into excited $K$-mesons
assuming that the \mbox{$s$-quark} may be treated in the static limit.
Thus we have to assign the states listed in
table~\ref{tab:K-resonances} to the
members of the spin symmetry doublets~(\ref{C}-\ref{G*}).
The assignment of the resonances to
spin symmetry doublets is not unique.
We shall choose our assignment of the excited $K$-mesons into the spin
symmetry doublets in the following way:
we put the $K_1 (1270)$ and the $K_2^* (1430)$ into the  spin
symmetry doublet $(E,E^*)$ and correspondingly the $K_1(1400)$ and
the $K(1430)$ into $(F,F^*)$.  In particular, this assignment of the
states $K_1(1270)$ and  $K_1(1400)$ means that due to spin symmetry
these will be a mixture of the quark model states $1^1P_1$ and $1^3P_1$
with a mixing angle of $45^\circ$.  In fact, this value of the mixing
angle is consistent with experimental results \cite{K1-mixing}, which
in turn supports our assignment of the states to the spin symmetry
doublets. A further motivation for our assignment is the fact that
in the $D$ system the lowest lying $1^+$ state is the spin symmetry
partner of the lowest $2^+$ state.  We shall put the
$K^*(1680)$ and the $K_2(1580)$ into a fourth spin symmetry doublet
$(G,G^*)$, assuming that they have both orbital angular momentum
$L=2$. Finally we assume that the $K^*(1460)$ and $K^*(1410)$ are
radially excited states, which form a fifth spin symmetry doublet
$(C,C^*)_2$.

Two more spin symmetry doublets could be formed by $K^*_0(1950)$,
$K_1(1650)$ $(E,E^*)_2$ and $K_2(1770)$, $K^*_3(1780)$ $(F,F^*)$.
However, we shall neither consider decays into radially excited states
(except for the low lying $(C,C^*)_2$ doublet), nor decays into states
with $J\ge3$.

%%%%%%%%%%%%%%%%%%%%%%%%%%%%%%%%%%%%%%%%%%%%%%%%%%%%%%%%%%%%%%%%%%%%%%%%

\section{Radiative Rare $B$ Decays into Excited $K$-Meson States}
\label{sec:penguins}

The radiative rare decays are mediated by the effective
Hamiltonian~\cite{effective Hamiltonian}
\begin{equation}
H_{eff} = - \frac{4 G_F}{\sqrt{2}} V_{tb} V^*_{ts} C_7 (m_b)
            {\cal O}_7 (m_b)
\end{equation}
where the operator ${\cal O}_7$ is given by
\begin{equation}
{\cal O}_7 = \frac{e}{32 \pi^2} F_{\mu \nu}
             (m_b \bar{s} \sigma^{\mu \nu} (1+\gamma_5) b
              + m_s \bar{s} \sigma^{\mu \nu} (1-\gamma_5) b )
\end{equation}
and the Wilson coefficient $C_7 (m_b)$ is
\begin{equation}
C_7 (m_b) = \eta^{-16/23} \left[ C_7 (M_W) + \frac{58}{135}
            \left(\eta^{10/23} - 1 \right)
            + \frac{29}{189}
            \left(\eta^{28/23} - 1 \right) \right]
\end{equation}
with $ \eta = \alpha_s (m_b) / \alpha_s (M_W) $ and
\begin{equation}
C_7 (M_W) = \frac{x}{2} \left[
            \frac{2 x^2 / 3 + 5 x / 12 - 7 / 12}{(x-1)^3}
          - \frac{3 x^2 / 2 - x}{(x-1)^4} \ln x \right],
          \quad x = \frac{m_t^2}{M_W^2}.
\end{equation}
The matrix element of ${\cal O}_7$ between a $B$-meson in the
initial state and a photon and a generic
$K^{**}$-meson in the final state may be written as
\begin{equation}
\ME{{\cal K}(v^\prime) , \gamma (q,\eta)}{{\cal O}_7}{B(v)}
= \frac{e}{16 \pi^2} \eta_\mu
\ME{{\cal K}(v^\prime) , \gamma (q,\eta)}{\Omega^\mu}{B(v)}
\end{equation}
where ${\cal K}$ denotes any of the six states
represented by~(\ref{C}-\ref{G*}),
\begin{equation}
\Omega_\mu = m_B \bar{s} \sigma_{\mu \nu} (1+\gamma_5) b q^\nu
          +  m_{K^{**}} \bar{s} \sigma_{\mu \nu} (1-\gamma_5) b q^\nu,
          \quad q = m_B v - m_{K^{**}} v^\prime,
\end{equation}
and $\eta_\mu$ is the polarization vector of the photon.

Using the mass shell condition of the photon
\begin{equation}
\label{eq:mass-shell}
  q^2 = m_B^2 + m_{K^{**}}^2 - 2 vv' m_B m_{K^{**}} = 0
\end{equation}
and the polarization sums for spin-$1$ and spin-$2$ particles
\begin{eqnarray}
  M_{\mu\nu}^{(1)} (v) & = & - g_{\mu\nu} + v_\mu v_\nu \\
  M_{\mu\nu,\rho\sigma}^{(2)} (v)
     & = & \frac{1}{2} M_{\mu\rho}^{(1)} (v) M_{\nu\sigma}^{(1)} (v)
       + \frac{1}{2} M_{\mu\sigma}^{(1)} (v) M_{\mu\sigma}^{(1)} (v)
       - \frac{1}{3} M_{\mu\nu}^{(1)} (v) M_{\rho\sigma}^{(1)} (v)  ,
\end{eqnarray}
we arrive at the following branching ratios:

\begin{eqnarray} \label{eq:br-first}
  \Gamma(B\rightarrow K^*\gamma)
     & = & \Omega \left\vert \xi_C (vv') \right\vert^2 \frac{1}{y}
              \left[ (1 - y)^2 (1 + y)^4 (1 + y^2) \right] \\
  \Gamma(B\rightarrow K_1(1270)\gamma)
     & = & \Omega \left\vert \xi_E (vv') \right\vert^2 \frac{1}{y}
              \left[ (1 - y)^4 (1 + y)^2 (1 + y^2) \right] \\
  \Gamma(B\rightarrow K_1(1400)\gamma)
     & = & \Omega \left\vert \xi_F (vv') \right\vert^2 \frac{1}{24y^3}
              \left[ (1 - y)^4 (1 + y)^6 (1 + y^2) \right] \\
  \Gamma(B\rightarrow K_2^*(1430)\gamma)
     & = & \Omega \left\vert \xi_F (vv') \right\vert^2 \frac{1}{8y^3}
              \left[ (1 - y)^4 (1 + y)^6 (1 + y^2) \right] \\
  \Gamma(B\rightarrow K^*(1680)\gamma)
     & = & \Omega \left\vert \xi_G (vv') \right\vert^2 \frac{1}{24y^3}
              \left[ (1 - y)^6 (1 + y)^4 (1 + y^2) \right] \\
  \Gamma(B\rightarrow K_2(1580)\gamma)
     & = & \Omega \left\vert \xi_G (vv') \right\vert^2 \frac{1}{8y^3}
              \left[ (1 - y)^6 (1 + y)^4 (1 + y^2) \right] \\
  \Gamma(B\rightarrow K^*(1410)\gamma)
     & = & \Omega \left\vert \xi_{C_2} (vv') \right\vert^2 \frac{1}{y}
              \left[ (1 - y)^2 (1 + y)^4 (1 + y^2) \right]
           \label{eq:br-last}
\end{eqnarray}

The argument $vv'$ of the Isgur-Wise functions is fixed by the
condition~(\ref{eq:mass-shell})
\begin{equation}
  vv' =  \frac{m_B^2 + m_{K^{**}}^2}{2 m_B m_{K^{**}}}
\end{equation}
and we have used the following abbreviations
\begin{equation}
  \Omega = \frac{\alpha}{128\pi^4} G_F^2 m_b^5
                \left\vert V_{tb} \right\vert^2
                \left\vert V_{ts} \right\vert^2
                \left\vert {\cal C}_7 (m_b) \right\vert^2, \quad
  y = \frac{m_{K^{**}}}{m_B}.
\end{equation}

Since the decays into states in the same spin symmetry doublet are
described by a single Isgur-Wise function, spin symmetry relates these
decays. From~(\ref{eq:br-first}-\ref{eq:br-last}) we find
\begin{eqnarray}
  \label{eq:factor-of-three}
  \Gamma(B\rightarrow K_2^*(1430)\gamma)
     & = & 3 \cdot \Gamma(B\rightarrow K_1(1400)\gamma) \\
  \Gamma(B\rightarrow K_2(1580)\gamma)
     & = & 3 \cdot \Gamma(B\rightarrow K^*(1680)\gamma) \nonumber
\end{eqnarray}
In the heavy quark limit the two members of a spin symmetry doublet
should be degenerate in mass, which means, that the corresponding
Isgur-Wise functions would be taken at the same value of $vv'$ and
the ratio $y$ would be the same for both states.
However, for the $s$-quark we expect a large breaking of the spin
symmetry, which means, that the two relations~(\ref{eq:factor-of-three})
will be only approximate. From the numerical estimates presented in
the next section one finds that the
relations~(\ref{eq:factor-of-three}) indeed hold within the indicated
accuracy.

%%%%%%%%%%%%%%%%%%%%%%%%%%%%%%%%%%%%%%%%%%%%%%%%%%%%%%%%%%%%%%%%%%%%%%%%
\pagebreak
\section{Wave Function Model for the Isgur-Wise\protect\goodbreak
         Functions}
\label{sec:wave-func}

Since we are dealing with two body decays, the product
$vv'$ is fixed by kinematics and hence the
Isgur-Wise functions have to be taken at this point.
Due to the large mass difference between the $B$
and the excited $K^{**}$-mesons, the $vv^\prime$ are
not close to unity and we shall use a model to
extrapolate from the point where the normalization of the usual
Isgur-Wise function is known.

We shall employ the wave function model of Grinstein, Scora, Isgur,
and Wise~\cite{GISW89} to determine the Isgur-Wise functions
in~(\ref{eq:br-first}-\ref{eq:br-last}).  In the context of this model
the Isgur-Wise
functions $\xi_{C,E,F,G}$ may be extracted from the following overlap
integral
\begin{equation}
\label{oi}
  I (\vec{v}^\prime ) = \int \frac{d^3 \vec{p}}{(2 \pi)^3 }
  \tilde \Phi_F^* (\vec{p} + \Lambda \vec{v}^\prime )
  \tilde \Phi_I (\vec{p})
  = \int d^3 \vec{x} \,
  \Phi_F^* (\vec{x}) \Phi_I (\vec{x})
  e^{-i \Lambda \vec{v}^\prime \cdot \vec{x}}
\end{equation}
where the labels $I$ and $F$ denote the wave function of the initial
and final meson respectively and the ``inertia parameter''
$\Lambda$ corresponds to the mass of
the light degrees of freedom.  In fact, since the $K$-mesons are not
particularly heavy, we shall use for $\Lambda$ the
expression~\cite{Alt88}
\begin{equation}
  \label{eq:lambda}
  \Lambda = \frac{m_{K^{**}} m_d}{m_s + m_d}
\end{equation}
which accounts for the kinematic effects of a finite $s$-quark mass.
The quark masses are taken to be
\begin{equation}
  m_d = 330\mbox{ MeV}, \qquad m_s = 550 \mbox{ MeV}.
\end{equation}

The wave functions are chosen to be eigenfunctions of orbital
angular momentum $L$, where
the initial meson will have $L = 0$; thus the wave functions are given by
\begin{equation}
  \label{eq:wavefun}
  \Phi_I (\vec x)
    = Y_0^0 \left(\frac{\vec x}{|\vec x|} \right) \phi_I (|\vec x|),
  \qquad
  \Phi_F (\vec x)
    = Y_L^m \left(\frac{\vec x}{|\vec x|} \right) \phi_F (|\vec x|)
\end{equation}
with the normalization
\begin{equation}
1 = \int d^3 \vec{x} \,
\Phi_{I/F}^* (\vec{x}) \Phi_{I/F} (\vec{x}) =
    \int dr \, r^2
\phi_{I/F}^* (r) \phi_{I/F} (r)
\end{equation}

Inserting the wave functions~(\ref{eq:wavefun})
into the overlap integral~(\ref{oi}) and
choosing the quantization axis of the orbital angular momentum in the
direction of the velocity $\vec v$, the overlap~(\ref{oi}) vanishes for
$m\neq0$. For $m=0$ the overlap integral becomes
\begin{equation} \label{master}
I(\vec{v'}) = \sqrt{2L+1}\, i^L \int r^2 dr \, \phi_F^* (r) \phi_I (r)
       j_L (\Lambda r\vert\vec{v'}\vert)
\end{equation}
where $j_L$ is the spherical Bessel function of order $L$.
This expression holds in the rest frame of the heavy meson;
in a general frame~(\ref{master}) becomes
\begin{equation}
  \label{eq:overlap}
  I(vv') = \sqrt{2L+1}\, i^L \int r^2 dr \, \phi_F^* (r) \phi_I (r)
       j_L (\Lambda r \sqrt{(vv')^2 - 1} )
\end{equation}
For $L=0$ the overlap integral is the usual
Isgur-Wise function, which is correctly normalized at $v \cdot
v^\prime = 1$, since $j_0 (0) = 1$. The overlap integrals for
$L = 1$ and $L = 2$ are identified with the remaining Isgur-Wise
functions $\xi_{E,F,G}$:
\begin{eqnarray}
\label{eq:xi*}
\xi_E (v \cdot v^\prime) = \xi_F (v \cdot v^\prime) &=&
\sqrt{3} i \int r^2 dr \, \phi_F^* (r) \phi_I (r)
       j_1 (\Lambda r \sqrt{(v \cdot v^\prime)^2 - 1} )  \\
\label{eq:xi***}
\xi_G (v \cdot v^\prime) &=&
- \sqrt{5}  \int r^2 dr \, \phi_F^* (r) \phi_I (r)
       j_2 (\Lambda r \sqrt{(v \cdot v^\prime)^2 - 1} )
\end{eqnarray}
Note that these functions will vanish at $v \cdot v^\prime = 1$ like
$((v \cdot v^\prime)^2 - 1)^{L/2}$, since the final state wave
function is orthogonal to the one of the initial state.

To calculate the overlap integrals we insert the radial wave functions
of the GISW model~\cite{GISW89} into eq.~(\ref{eq:overlap}).  These
correspond to harmonic
oscillator wave functions with oscillator strengths $\beta_K$ and
$\beta_B$.  The values fitted in~\cite{GISW89} to the semileptonic $B$ and
$D$ decays ($\beta_K = 0.34\mbox{ GeV}$, $\beta_B = 0.41\mbox{ GeV}$)
are not equal.  This breaking of the heavy flavor symmetry causes a
violation of the normalization condition $\xi_C(1) = 1$.
To estimate the model dependence and the effect of the breaking of
heavy quark symmetry we use the same parameter $\beta$ for both wave
functions and vary it between $\beta_K$ and $\beta_B$. The resulting
ranges for the Isgur-Wise functions are shown in the shaded bands of
figure~\ref{fig:iwf}.  For comparison, we have also plotted the
Isgur-Wise function $\xi_C$ as calculated in~\cite{NR91} from the
model~\cite{BSW85}. In the region of interest for the decays
$B\to K^{*}\gamma$, $B\to K^{**}\gamma$ discussed in this paper, the
variable $vv'$ lies in the range $1.70 \le vv' \le 3.05$.  We
see from figure~\ref{fig:iwf} that the Isgur-Wise functions $\xi_C$
determined from the two models are very similar in this range,
particularly for $vv' \ge 2$.

%%% if EPSF %%%%%%%%%%%%%%%%%%%%%%%%%%%%%%%%%%%%%%%%%%%%%%%%%%%%%%%%%%%%
\begin{figure}
  \leavevmode
  \begin{center}
    \epsfxsize=\linewidth
    \leavevmode
    \epsffile{iwf.eps}
  \end{center}
  \caption{Plot of the Isgur-Wise functions $\xi_C, \xi_{E,F}, \xi_G$
    as a function of the variable $vv'$.
    The shaded bands give the range of our estimate as obtained from
    the model~\protect\cite{GISW89} by varying the parameter
    $\beta$ as discussed in the text.  For comparison, the Isgur-Wise
    function $\xi_C$ calculated from the model~\protect\cite{BSW85}
    is shown by the solid curve.}
  \label{fig:iwf}
\end{figure}
%%% fi %%%%%%%%%%%%%%%%%%%%%%%%%%%%%%%%%%%%%%%%%%%%%%%%%%%%%%%%%%%%%%%%%

Table~\ref{tab:results} shows our estimates for the exclusive $B\to
K^{**}\gamma$ branching ratios.  Our results are given in units of the
inclusive branching ratio $B\to X_s\gamma$, which is estimated by the
QCD improved $b \to s\gamma$ quark decay rate
\begin{equation}
  \Gamma(B\to X_s\gamma) \approx \Gamma(b\to s\gamma) =
         4 \Omega (1 - y)^3 (1 + y)^3 (1 + y^2),
\end{equation}
which corresponds to an inclusive branching ratio
$BR(B\to X_s\gamma) = (3 - 5) \times 10^{-4}$~\cite{AG91}.  For the
numerical values of the $B\to K^{**}\gamma$ branching ratios in
table~\ref{tab:results} we have used
$BR(B\to X_s\gamma) = 4 \times 10^{-4}$, corresponding to a top quark
mass of about 150 GeV.  The ranges are obtained from the variation of
the model parameters as mentioned above.

\begin{table}
  \begin{center}
  \begin{tabular}{|l|l|c|c|r@{ -- }r|r@{ -- }r|r@{ -- }r|}
  \hline
    \multicolumn{1}{|c|}{Name}
       &\multicolumn{1}{|c|}{State}
       &$ J^P $&$ vv' $
       &\multicolumn{2}{|c|}{$\xi$}
       &\multicolumn{2}{|c|}{$R$}
       &\multicolumn{2}{|c|}{$BR \times 10^5$} \\
  \hline \hline
    $ K           $&$ C     $&$ 0^- $& 5.346 &  0.125 &  0.239
       & \multicolumn{2}{|c|}{forbidden}
       & \multicolumn{2}{|c|}{forbidden} \\
    $ K^*         $&$ C^*   $&$ 1^- $& 3.030 &  0.136 &  0.253
       &  3.5\% & 12.2\% &  1.4 &  4.9 \\
    $ K^*(1430)   $&$ E     $&$ 0^+ $& 1.982 &  0.309 &  0.453
       & \multicolumn{2}{|c|}{forbidden}
       & \multicolumn{2}{|c|}{forbidden} \\
    $ K_1(1270)   $&$ E^*   $&$ 1^+ $& 2.198 &  0.296 &  0.441
       &  4.5\% & 10.1\% &  1.8 &  4.0 \\
    $ K_1(1400)   $&$ F     $&$ 1^+ $& 2.015 &  0.307 &  0.451
       &  6.0\% & 13.0\% &  2.4 &  5.2 \\
    $ K^*_2(1430) $&$ F^*   $&$ 2^+ $& 1.987 &  0.309 &  0.452
       & 17.3\% & 37.1\% &  6.9 & 14.8 \\
    $ K^*(1680)   $&$ G     $&$ 1^- $& 1.702 &  0.359 &  0.420
       &  1.1\% &  1.5\% &  0.4 &  0.6 \\
    $ K_2(1580)   $&$ G^*   $&$ 2^- $& 1.820 &  0.350 &  0.417
       &  4.5\% &  6.4\% &  1.8 &  2.6 \\
    $ K(1460)     $&$ C_2   $&$ 0^- $& 1.946 & -0.242 & -0.293
       & \multicolumn{2}{|c|}{forbidden}
       & \multicolumn{2}{|c|}{forbidden} \\
    $ K^*(1410)   $&$ C^*_2 $&$ 1^- $& 2.005 & -0.240 & -0.292
       &  7.2\% & 10.6\% &  2.9 &  4.2 \\
  \hline\hline
    Sum            &         &       &       & \multicolumn{2}{|c|}{}
       & 44.1\% & 90.9\% & 17.6 & 36.4 \\
  \hline
  \end{tabular}
  \end{center}
\caption{Values of the Isgur-Wise function at the indicated value of
         $vv'$, the ratio $R = \Gamma(B\to K^{**}\gamma)/\Gamma(B\to
         X_s\gamma)$, and the branching ratio $BR(B\to K^{**}\gamma)$
         for the various excited $K^{**}$-mesons with the indicated
         mass and $J^P$.}
\label{tab:results}
\end{table}

To compare the numbers in table~\ref{tab:results} with the results
of~\cite{Alt88}, we have to note first that in the meantime
the experimental information
on the $K^{**}$-resonances has increased considerably.  Taking into
account updated mass values, our results are generally in agreement
with~\cite{Alt88}.  However, there is a noticeable difference in our
prediction for the branching fraction into $K^*_2(1430)$, which
is from our estimate about a factor of three larger compared
to~\cite{Alt88}.  Our result is consistent with spin symmetry,
because it satisfies the general relations~(\ref{eq:factor-of-three})
which hold without model dependent input for the Isgur-Wise
functions.

In quark model calculations~\cite{Alt88}, the decay into the $1^1P_1$
state ($K_{1B}$) is forbidden, because ${\cal O}_7$ is a spin-flip
operator.  In our analysis both $K_1(1270)$ and $K_1(1400)$
contribute since they are states with
good angular momentum of the light degrees of freedom
and thus they both contain a $1^3P_1$ component.
The two channels $K^*(892)$ and $K^*_2(1430)$ are the easiest channels
to observe, since the efficiency for the reconstruction of the decaying
$K^*$-meson is high in both cases~\cite{Alb89}.
In particular the $K^*_2(1430)$ is a
very interesting candidate, since our model calculation yields a
very high branching fraction into this channel.

%%%%%%%%%%%%%%%%%%%%%%%%%%%%%%%%%%%%%%%%%%%%%%%%%%%%%%%%%%%%%%%%%%%%%%%%

\section{Conclusions}
\label{sec:concl}

We have studied the predictions of the heavy quark symmetry for the rare
decays of $B$-mesons into higher $K$-resonances $B\to K^{**}\gamma$.
In contrast to
earlier calculation, we find that a substantial fraction (17-37\%)
of the inclusive $b\to s\gamma$ branching ratio goes into the
$K^*_2(1430)$ channel, which should be relatively easy to observe.

Not much is known about the Isgur-Wise functions for the different
spin symmetry doublets.  To obtain more
information one needs to use model dependent input. We have employed a
specific model~\cite{GISW89} to calculate the Isgur-Wise functions for
the various spin symmetry doublets.  To estimate the uncertainties
introduced by the model we have varied the model parameters within a
range corresponding to the breaking of heavy quark symmetry.  In
defense of the GISW model, we can add that its predictions for the
$D^*$ and higher resonances in the semileptonic $B$ decays
$B\to D^*\ell\nu_\ell$ and $B\to D^{**}\ell\nu_\ell$ are in reasonable
agreement with recent data~\cite{Alb92,Sto92}.

Since the decays studied here are all two body decays, the corresponding
Isgur-Wise functions are taken at a single point, as $vv^\prime$
is fixed by kinematics.
Due to the large mass difference between the $B$-meson and all the $K$
resonances the values for $vv^\prime$ range between 2 and 3 and thus a
considerably large extrapolation off the normalization point $vv^\prime=1$
is necessary.  It remains to be tested whether the representation of
the Isgur-Wise functions assumed here is trustworthy.  Radiative rare
$B$ decays $B\to K^*\gamma$, $B\to K^{**}\gamma$ provide interesting
avenues for checking the predictions of the heavy quark symmetry
combined with meson wave function models.

We expect that the estimates presented in this paper should hold
within the indicated range; in particular we expect a relatively large
branching fraction for the channel $K^*_2(1430)$ which is according to our
estimate a factor of 3 to 4 larger than the one for the $K^*(892)$.
The resulting hadron mass spectra are also in qualitative agreement
with the estimates for the inclusive radiative $B$ decays
$B\to X_s\gamma$~\cite{AG91}.

%%%%%%%%%%%%%%%%%%%%%%%%%%%%%%%%%%%%%%%%%%%%%%%%%%%%%%%%%%%%%%%%%%%%%%%%

%%%%%%%%%%%%%%%%%%%%%%%%%%%%%%%%%%%%%%%%%%%%%%%%%%%%%%%%%%%%%%%%%%%%%%%%

\end{document}